\documentclass[pre,aps,preprint,showpacs,preprintnumbers,amsmath,amssymb,secnumroman,eqsecnum]{revtex4}

\usepackage{graphicx}
\usepackage{dcolumn}
\usepackage{bm}

\newcommand{\beq}{\begin{equation}}
\newcommand{\eeq}{\end{equation}}
\newcommand{\beqa}{\begin{eqnarray}}
\newcommand{\eeqa}{\end{eqnarray}}
\newcommand{\vc}[1]{\mbox{\boldmath $#1$}}

\newcommand{\vol}[1]{{\bf #1}}

\begin{document}


\title{Vanishing mean volume velocity in isothermal isobaric diffusion of a binary fluid mixture}

\author{B. U. Felderhof}

 \email{ufelder@physik.rwth-aachen.de}
\affiliation{Institut f\"ur Theorie der Statistischen Physik \\ RWTH Aachen University\\
Templergraben 55\\52056 Aachen\\ Germany\\
}%

\date{\today}

\begin{abstract}
It is shown that in isothermal isobaric diffusion of a binary fluid mixture the mean volume velocity vanishes in the linear regime, independent of the equation of state. The partial specific volumes of the two components are uniform and constant in the process of mutual diffusion. The properties lead to a simple derivation of the de Groot-Mazur thermodynamic factor in the diffusion coefficient. The properties also imply that the diffusive volume flux defined by Brenner is proportional to the mass current density, and is therefore not a quantity of independent interest.
\end{abstract}

\pacs{05.60.Cd, 5.70.Ln, 47.57.eb, 82.60.Lf}
\maketitle
\section{\label{I}Introduction}

The value of the mean volume velocity in a diffusing binary fluid mixture has been a subject of much debate.
In their monograph on non-equilibrium thermodynamics de Groot and Mazur \cite{1} stated that the mean volume
velocity vanishes in systems where the partial specific volumes do not depend appreciably on the concentrations
and the pressure. We show in the following that in linear irreversible thermodynamics the condition can be dropped. It follows from the condition of
mechanical equilibrium that in the absence of applied potentials the pressure is uniform. This is sufficient to
show that in the linear diffusive regime the mean volume velocity vanishes and that the two partial specific volumes are
uniform and constant in time.

In the first section of their treatise on multicomponent mass transfer \cite{2} Taylor and Krishna state that the literature on diffusion would be a good deal simpler if there were only one way to define diffusion fluxes. The theorem derived below shows that at least there is a preferred reference frame, namely the laboratory frame with the diffusion cell at rest and with vanishing mean volume velocity. As we remarked earlier \cite{3}, the transformation to a different frame has only local significance. Much effort was spent on such transformations and the corresponding definition of diffusion coefficients \cite{1},\cite{4},\cite{5}. In our view it is preferable to consider from the start the laboratory frame with the cell at rest.

In the theoretical analysis we consider a binary fluid mixture which initially at time $t=0$ is in a non-uniform non-equilibrium state.
The containing vessel is closed to particle exchange, but is in
thermal contact with a thermal reservoir at constant temperature
$T$. In the course of time the system
relaxes to the final uniform equilibrium
state. De Groot and Mazur \cite{1} argued that in most liquid
mixtures there is a separation of time scales such that the
pressure and temperature equilibrate first, and become uniform before the
composition of the system changes appreciably. Subsequently the
system remains in mechanical equilibrium at the same uniform
pressure and temperature, and particles diffuse till also chemical equilibrium is
reached \cite{6}.

As discussed by van Kampen \cite{7}, quite generally in systems
with a wide separation of time scales the slow regime following
the fast initial stage can be described by a reduced number of
variables. In the case of a binary fluid mixture the variables in
the diffusive regime are the two number densities. The momentum
densities, the pressure, and the temperature can be ignored. The initial stage
must be described by hydrodynamic equations and the equation of
thermal conduction, but these need not be discussed here.

In the subsequent slow diffusive regime the system is isothermal
and isobaric. Near equilibrium the deviations of the two particle number densities from their equilibrium values are proportional
to each other and obey the same linear diffusion equation. In Sec. II we prove that in the linear regime the two partial specific volumes are uniform and constant, and that the mean volume velocity vanishes. In Sec. III we use the latter property for a simple derivation of the de Groot-Mazur thermodynamic factor occurring in the diffusion coefficient.

Brenner discussed the transport of volume in diffusion in a series
of articles \cite{8}-\cite{12}. He claimed that his findings add
credibility to his theory of bivelocity hydrodynamics for both
gaseous and liquid continua. In Sec. IV we show that the
diffusive volume flux defined by Brenner \cite{10},\cite{11} is
directly proportional to the mass current density. Therefore it is not
a quantity of independent interest. The actual volume flux vanishes.

Bringuier \cite{13} performed a singular perturbation
analysis of the hydrodynamic equations, and came to the conclusion
that in interdiffusion the mass transport velocity vanishes, but
that the volume velocity does not vanish. He claimed that this has
strong implications for the measurement of mutual diffusivities in
non-dilute mixtures. We argue below that he ignores that in the
diffusive regime the barycentric velocity is no longer a relevant
variable. His conclusion does not hold.

\section{\label{II}Mechanical equilibrium and volume flux}

We consider a binary fluid mixture of $N_0$ solvent particles and $N_1$ solute particles enclosed in a vessel of volume $V$. We assume that there are no applied potentials, so that in the final thermal equilibrium state the system is uniform at temperature $T$ with number densities $n_{0eq}=N_0/V$ and $n_{1eq}=N_1/V$. In the equilibrium state the pressure is also uniform with value $p_{eq}$.

In non-equilibrium thermodynamics the system is described in a continuum picture with a local mass density $\rho(\vc{r},t)$, a barycentric velocity $\vc{v}(\vc{r},t)$, number densities $n_0(\vc{r},t)$ and $n_1(\vc{r},t)$, temperature $T(\vc{r},t)$, pressure $p(\vc{r},t)$, and entropy density $s(\vc{r},t)$. The system is assumed to be locally in equilibrium, so that the pressure and entropy density are given in terms of number densities and temperature by equations of state, which are assumed to be known.

The approach to equilibrium is governed by transport equations which we do not specify in detail. Instead we shall analyze the approach to equilibrium in terms of a simplified picture, first developed by de Groot and Mazur \cite{1}, in which it is assumed that there is a separation of time scales. From a given non-equilibrium state there is first a fast process of flow and heat conduction after which the pressure and temperature are uniform, but the particle densities are still non-uniform. The fast process is followed by a slow process of isothermal isobaric diffusion in which the particle densities finally tend to their uniform equilibrium values $n_{0eq}$ and $n_{1eq}$. We call the isothermal isobaric state reached at time $t_i$ after the initial fast process the interim state.

In the interim state the system is in mechanical equilibrium and the pressure and temperature are uniform \cite{1}, \cite{2}. At each point $\vc{r}$ and for all $t\geq t_i$
\begin{equation}
\label{2.1}p(n_0(\vc{r},t),n_1(\vc{r},t),T)=p(n_{0eq},n_{1eq},T)=p_{eq}.
\end{equation}
The partial specific volumes $\overline{v}_0$ and $\overline{v}_1$ are given by \cite{3},\cite{14}
\begin{equation}
\label{2.2}\overline{v_j}=\frac{\partial p/\partial n_j}{n_0\partial p/\partial n_0+n_1\partial p/\partial n_1},\qquad (j=0,1).
\end{equation}
These quantities vary along the isobar, but satisfy
\begin{equation}
\label{2.3}n_0\overline{v_0}+n_1\overline{v_1}=1.
\end{equation}
The expression Eq. (2.2) agrees with that of Kirkwood and Buff \cite{15} in terms of correlation integrals \cite{16}.

Differentiating Eq. (2.1) with respect to position and using Eq. (2.3) we find
\begin{equation}
\label{2.4}n_0\nabla\overline{v_0}+n_1\nabla\overline{v_1}=0.
\end{equation}
Differentiating Eq. (2.1) with respect to time and using Eq. (2.3)
we find
\begin{equation}
\label{2.5}n_0\frac{\partial\overline{v_0}}{\partial t}+n_1\frac{\partial\overline{v_1}}{\partial t}=0.
\end{equation}
Eqs. (2.4) and (2.5) hold for all $\vc{r}$ in the vessel and for
all $t\geq t_i$.

We show that in linear irreversible thermodynamics the partial specific
volumes $\overline{v_0},\overline{v_1}$ are uniform, constant, and
equal to the specific volumes $\overline{v_0}_{eq}$ and
$\overline{v_1}_{eq}$ of the final equilibrium state. To first order in the deviations from equilibrium $n_0^1,n_1^1$, defined by
\begin{equation}
\label{2.6}n_0(\vc{r},t)=n_{0eq}+n_0^1(\vc{r},t),\qquad n_1(\vc{r},t)=n_{1eq}+n_1^1(\vc{r},t),
\end{equation}
the condition of  mechanical equilibrium Eq. (2.1) reads
\begin{equation}
\label{2.7}n_0^1(\vc{r},t)\overline{v_0}_{eq}+n_1^1(\vc{r},t)\overline{v_1}_{eq}=0,
\end{equation}
where the coefficients $\overline{v_0}_{eq},\overline{v_1}_{eq}$ are uniform and constant. This shows that to first order $\overline{v_0},\overline{v_1}$ in Eq. (2.3) can be replaced by $\overline{v_0}_{eq},\overline{v_1}_{eq}$. In linear irreversible thermodynamics the pressure can be approximated by the linear expression
\begin{equation}
\label{2.8}P=p_{eq}+A_0(n_0-n_{0eq})+A_1(n_1-n_{1eq}),
\end{equation}
where the coefficients $A_0,A_1$ follow from the Taylor expansion of $p(n_0,n_1,T)$ about $n_{0eq},n_{1eq}$. In this approximation the partial specific volumes found from Eq. (2.2) are
\begin{equation}
\label{2.9}\overline{v_0}_{P}=\frac{A_0}{A_0n_{0}+A_1n_{1}},\qquad \overline{v_1}_{P}=\frac{A_1}{A_0n_{0}+A_1n_{1}}.
\end{equation}
Along the isobar $P=p_{eq}$ we have
\begin{equation}
\label{2.10}A_0n_0+A_1n_1=A_0n_{0eq}+A_1n_{1eq},
\end{equation}
and correspondingly
\begin{equation}
\label{2.11}\overline{v_0}_{P}=\overline{v_0}_{eq},\qquad\overline{v_1}_{P}=\overline{v_1}_{eq},
\end{equation}
are constant, confirming the statement before Eq. (2.6).

The number densities $n_0(\vc{r},t_i)$ and $n_1(\vc{r},t_i)$ serve as initial values for the diffusion equations describing isothermal isobaric diffusion in the approach to the final equilibrium state. In the linear regime the phenomenological equations take the form
\begin{equation}
\label{2.12}\frac{\partial n_0}{\partial t}=D\nabla^2 n_0,\qquad\frac{\partial n_1}{\partial t}=D\nabla^2 n_1,
\end{equation}
with the same diffusion coefficient $D$ in both equations.

The balance equations for the two number densities read in general
\begin{equation}
\label{2.13}\frac{\partial n_0}{\partial t}+\nabla\cdot\vc{j}_0=0,
\qquad\frac{\partial n_1}{\partial t}+\nabla\cdot\vc{j}_1=0.
\end{equation}
Correspondingly we define the mean volume velocity $\vc{v}^0$ as \cite{1},\cite{17}
\begin{equation}
\label{2.14}\vc{v}^0=\overline{v_0}\vc{j}_0+\overline{v_1}\vc{j}_1.
\end{equation}
In the linear regime $\overline{v_0}=\overline{v_0}_{eq},\overline{v_1}=\overline{v_1}_{eq}$ for states along the isobar. Then we find with $\vc{j}_0=-D\nabla n_0,\;\vc{j}_1=-D\nabla n_1$ from Eq. (2.7) that $\vc{v}^0$ vanishes
for all $\vc{r}$ in the vessel.

The derivation shows that the absence of volume flux follows in the linear regime from the condition of mechanical equilibrium. Onsager \cite{17} required that in diffusion the bulk displacement of the liquid vanishes and took this to imply that  $\vc{v}^0=0$. Katchalsky and Curran \cite{18} took the property to be self-evident, and simply stated that in ordinary diffusion the total volume flux is zero. A similar statement is made by Berne and Pecora \cite{19}. The property was denied by Bringuier \cite{13}, who claimed that instead the mass transport velocity vanishes.

\section{\label{III}Isothermal isobaric diffusion}

After time $t_i$ the pressure remains uniform and constant. From the Gibbs-Duhem relation we therefore find that the gradients of number densities balance as
\begin{equation}
\label{3.1}n_0\vc{X}_0+n_1\vc{X}_1=0,
\end{equation}
with thermodynamic forces $\vc{X}_0,\;\vc{X}_1$ given by
 \begin{eqnarray}
\label{3.2}\vc{X}_0&=&-\frac{\partial \mu_0}{\partial n_0}\nabla n_0-\frac{\partial \mu_0}{\partial n_1}\nabla n_1\nonumber\\
\vc{X}_1&=&-\frac{\partial \mu_1}{\partial n_0}\nabla n_0-\frac{\partial \mu_1}{\partial n_1}\nabla n_1.
\end{eqnarray}
The expressions for the two chemical potentials $\mu_0(n_0,n_1,T),\;\mu_1(n_0,n_1,T)$ are derived from that for the Helmholtz free energy.

We noted below Eq. (2.14) that in linear irreversible thermodynamics the mean volume velocity vanishes,
\begin{equation}
\label{3.3}\overline{v_0}_{eq}\vc{j}_0+\overline{v_1}_{eq}\vc{j}_1=0.
\end{equation}
With local entropy production $\sigma$ defined by
\begin{equation}
\label{3.4}T\sigma=\vc{j}_0\cdot\vc{X}_0+\vc{j}_1\cdot\vc{X}_1,
\end{equation}
we find by use of Eqs. (2.4), (3.1), and (3.3) in linear irreversible thermodynamics
\begin{equation}
\label{3.5}T\sigma=\frac{1}{n_{0eq}\overline{v}_{0eq}}\;\vc{j}_1\cdot\vc{X}_1.
\end{equation}
The formalism of linear irreversible thermodynamics \cite{1} leads to the relation
\begin{equation}
\label{3.6}\vc{j}_1=\frac{L_1}{n_{0eq}\overline{v}_{0eq}}\vc{X}_1,
\end{equation}
with positive Onsager coefficient $L_1$. By use of Eqs. (2.2) and (2.7) we can write
\begin{equation}
\label{3.7}\vc{X}_1=-\bigg[\frac{\partial \mu_1}{\partial n_1}-\frac{\partial \mu_1}{\partial n_0}\frac{\partial p/\partial n_1}{\partial p/\partial n_0}\bigg]\nabla n_1
=-\bigg(\frac{\partial\mu_1}{\partial n_1}\bigg)_p\nabla n_1.
\end{equation}
The relation Eq. (3.6) implies the linear diffusion equation
\begin{equation}
\label{3.8}\frac{\partial n^1_1}{\partial t}=D\nabla^2n^1_1,
\end{equation}
with diffusion coefficient
\begin{equation}
\label{3.9}D=\frac{L_1}{n_{0eq}\overline{v}_{0eq}}\bigg(\frac{\partial\mu_1}{\partial n_1}\bigg)_p\bigg|_{eq}.
\end{equation}
The number density $n^1_0$ satisfies the same equation on account of Eq. (2.7). With
\begin{equation}
\label{3.10}L_1=\frac{n_{1eq}}{\zeta^*_{10}},
\end{equation}
where $\zeta^*_{10}$ is the friction coefficient, the diffusion coefficient takes the form \cite{3},\cite{16}
\begin{equation}
\label{3.11}D=\frac{Q_1}{\zeta^*_{10}}\bigg|_{eq},\qquad Q_1=\frac{n_{1}}{n_{0}\overline{v}_{0}}\bigg(\frac{\partial\mu_1}{\partial n_1}\bigg)_p,
\end{equation}
where $Q_1$ is the de Groot-Mazur thermodynamic factor \cite{16}.

\section{\label{IV}Diffusive volume flux and mass flux}

Some years ago Brenner \cite{10},\cite{11} claimed to present evidence in support of the existence of a diffusive flux of volume in mixtures. He regarded this as a critical test of bivelocity hydrodynamics for mixtures. We show below that the diffusive volume flux is simply proportional to the mass current density.

The mass current density of the binary fluid mixture is defined as
\begin{equation}
\label{4.1}\vc{j}_m=m_0\vc{j}_0+m_1\vc{j}_1,
\end{equation}
where $m_0, m_1$ is the mass of a particle of species $0,1$. The individual components are
\begin{equation}
\label{4.2}\vc{j}_{m0}=m_0\vc{j}_0,\qquad \vc{j}_{m1}=m_1\vc{j}_1.
\end{equation}
Brenner \cite{11} defines diffusive mass fluxes $\vc{d}_{m0},\vc{d}_{m1}$ (in modified notation) by
\begin{equation}
\label{4.3}\vc{j}_{m0}=w_{0m}\vc{j}_m+\vc{d}_{m0},\qquad \vc{j}_{m1}=w_{1m}\vc{j}_m+\vc{d}_{m1},
\end{equation}
where the weights $w_{0m},w_{1m}$ are the mass fractions
\begin{equation}
\label{4.4}w_{0m}=\frac{n_0m_0}{n_0m_0+n_1m_1},\qquad w_{1m}=\frac{n_1m_1}{n_0m_0+n_1m_1}.
\end{equation}
The diffusive mass fluxes are related to the current densities by
\begin{equation}
\label{4.5}\vc{d}_{m0}=\frac{m_0m_1}{n_0m_0+n_1m_1}\;(n_1\vc{j}_0-n_0\vc{j}_1),\qquad \vc{d}_{m1}=-\vc{d}_{m0}.
\end{equation}
Brenner \cite{11} defines the diffusive volume flux $\vc{j}_v$ as
\begin{equation}
\label{4.6}\vc{j}_v=\overline{v_{0m}}\vc{d}_{m0}+\overline{v_{1m}}\vc{d}_{m1}=(\overline{v_{1m}}-\overline{v_{0m}})\vc{d}_{m1},
\end{equation}
with partial specific volumes per unit mass defined as
\begin{equation}
\label{4.7}\overline{v_{0m}}=\frac{\overline{v_0}}{m_0},\qquad\overline{v_{1m}}=\frac{\overline{v_1}}{m_1},
\end{equation}
(again in modified notation). If one writes \cite{8},\cite{9}
\begin{equation}
\label{4.8}\vc{j}_v=L_{v1}\vc{X}_1,\qquad\vc{d}_{m1}=L_{11}\vc{X}_1,
\end{equation}
then one has from Eq. (4.6)
\begin{equation}
\label{4.9}L_{v1}=\bigg(\frac{\overline{v_{1eq}}}{m_1}-\frac{\overline{v_{0eq}}}{m_0}\bigg)L_{11},
\end{equation}
since in linear irreversible thermodynamics the densities $n_0,n_1$ in Eq. (4.7) can be replaced by $n_{0eq},n_{1eq}$.
Brenner \cite{11} regards Eq. (4.9) as an important relation which can be tested. However, it follows simply from the definitions. In the ideal gas limit $\overline{v_{0}}=\overline{v_{1}}=1/(n_0+n_1)$, and we recover Brenner's \cite{11} Eq. (5.4).

By use of the above definitions and the relation $\overline{v_{0}}\vc{j}_0+\overline{v_{1}}\vc{j}_1=0$ the diffusive volume flux can be rewritten as
\begin{equation}
\label{4.10}\vc{j}_v=C_{10}\vc{j}_1,
\end{equation}
with coefficient
\begin{equation}
\label{4.11}C_{10}=\frac{1}{\overline{v_{0}}}\;\frac{m_0\overline{v_{1}}-m_1\overline{v_{0}}}{n_0m_0+n_1m_1}.
\end{equation}
In the linear regime the partial volumes
$\overline{v_{0}},\overline{v_{1}}$ are constant and can be
replaced by $\overline{v_{0}}_{eq},\overline{v_{1}}_{eq}$. The
corresponding $C_{10eq}$ is a constant. As shown in Eq.
(4.10), in the linear theory the so-called diffusive volume flux
$\vc{j}_v(\vc{r},t)$ is proportional to the particle current
density $\vc{j}_1(\vc{r},t)$ and has no physical significance of
its own  beyond its definition Eq. (4.6). The actual volume flux
$\vc{v}^0$ vanishes.

The mass current density $\vc{j}_m$ defined in Eq. (4.1) occurs in the mass continuity equation
\begin{equation}
\label{4.12}\frac{\partial\rho}{\partial t}+\nabla\cdot\vc{j}_m=0,
\end{equation}
where $\rho=n_0m_0+n_1m_1$ is the local mass density. In diffusion the mass current density is also just proportional to the particle current density $\vc{j}_1$, since
\begin{equation}
\label{4.13}\vc{j}_m=(m_1-\frac{\overline{v_{1}}}{\overline{v_{0}}}m_0)\vc{j}_1.
\end{equation}
By comparison with Eqs. (4.10) and (4.11) we find
\begin{equation}
\label{4.14}\vc{j}_m=-\rho\vc{j}_v.
\end{equation}
In the linear regime the mass density $\rho$ can be replaced by the constant factor $\rho_{eq}$. This shows that the diffusive volume flux $\vc{j}_v$ is not a quantity of independent interest.

We conclude that the diffusive volume flux $\vc{j}_v(\vc{r},t)$ and the mass flux $\vc{j}_m(\vc{r},t)$ are just the particle current density $\vc{j}_1(\vc{r},t)$ in disguise. No separate phenomenological laws are needed for these quantities. The two particle current densities $\vc{j}_0(\vc{r},t)$ and $\vc{j}_1(\vc{r},t)$ are simply proportional by the relation Eq. (3.3), which expresses the absence of volume flow. The number densities $n_0(\vc{r},t)$ and $n_1(\vc{r},t)$ vary as given by Eqs. (2.7) and (2.12). The fact that in the diffusive regime the mean volume velocity $\vc{v}^0$ vanishes, but $\vc{j}_m$ differs from zero, cannot be taken as support for the validity of bivelocity hydrodynamics, since in this regime hydrodynamics is not relevant.

In the diffusive regime only the number densities $n_0,n_1$ and the corresponding current densities $\vc{j}_0,\vc{j}_1$ are relevant. It does not make sense to write $\vc{j}_m=\rho\vc{v}$, as done by Bringuier \cite{13}. The barycentric velocity $\vc{v}$ thus defined does not satisfy an independent hydrodynamic equation of motion. Bringuier \cite{13} claimed that to lowest order in his expansion scheme $\vc{v}=0$, and that correspondingly the volume current density does not vanish in the diffusive regime. This is in conflict with the condition of mechanical equilibrium Eq. (2.1). In effect Bringuier denies that the diffusion process is isobaric.

\section{\label{V}Discussion}
We showed above that the condition of mechanical equilibrium Eq. (2.1) in conjunction with the expression Eq. (2.2) for the partial specific volumes can be used to prove that the diffusion of a binary mixture in the linear regime has the simple property that the partial specific volumes of both components are uniform and constant, and that the mean volume velocity vanishes. The proof is important, since it shows that the volume-fixed frame is to be preferred in theoretical analysis, as well as in computer simulation. As noted by Fitts \cite{20}, the volume-fixed frame is the one preferred by experimentalists. He argued also that this frame has the advantage that the diffusion coefficients of the two components are equal. In centrifugal analysis the pressure does vary in the stationary state, and the corresponding variation of the partial specific volumes may play a role \cite{21}.

In isothermal isobaric diffusion of a binary mixture the deviations of the two particle number densities from their equilibrium values are simply proportional, as shown in Eq. (2.7). Correspondingly, in the linear regime the mean volume velocity, defined in Eq. (2.14), vanishes everywhere. In Sec. IV we used this property to show that Brenner's concept of a diffusive volume flux is not pertinent.


\newpage


\begin{thebibliography}{99}

\bibitem{1}
S. R. de Groot and P. Mazur, {\it Non-equilibrium thermodynamics} (North-Holland, Amsterdam, 1962).

\bibitem{2}
R. Taylor and R. Krishna, {\it Multicomponent mass trasnfer} (Wiley, New York, 1993).

\bibitem{3}
B. U. Felderhof, J. Chem. Phys. \vol{118}, 11326 (2003).

\bibitem{4}
J. G. Kirkwood, R. L. Baldwin, P. J. Dunlop, L. J. Gosling, and G. Kegeles, J. Chem. Phys. \vol{33}, 1505 (1960).

\bibitem{5}
H. M. J. Boots and J. M. Deutch, Physica A \vol{94}, 99 (1978).

\bibitem{6}
J. G. Kirkwood and I. Oppenheim, {\it Chemical Thermodynamics} (McGraw-Hill, New York, 1961).

\bibitem{7}
N. G. van Kampen, Phys. Rep. \vol{124}, 69 (1985).

\bibitem{8}
H. Brenner, Physica A \vol{349}, 59 (2005).

\bibitem{9}
H. Brenner, Int. J. Eng. Sci. \vol{47}, 902 (2009).

\bibitem{10}
H. Brenner, J. Chem. Phys. \vol{132}, 054106 (2010).

\bibitem{11}
H. Brenner, J. Chem. Phys. \vol{133}, 154102 (2010).

\bibitem{12}
H. Brenner, Ind. Eng. Chem. Res. \vol{50}, 8927 (2011).

\bibitem{13}
E. Bringuier, Physica A \vol{391}, 5064 (2012).

\bibitem{14}
S. Vafaei, B. Tomberli, and C. G. Gray, J. Chem. Phys. \vol{141}, 154501 (2014).

\bibitem{15}
J. G. Kirkwood and F. P. Buff, J. Chem. Phys. \vol{19}, 774 (1951).

\bibitem{16}
B. U. Felderhof, J. Chem. Phys. \vol{147}, 074902 (2017).

\bibitem{17}
L. Onsager, Ann. N. Y. Acad. Sci. \vol{46}, 241 (1945).

\bibitem{18}
A. Katchalsky and P. F. Curran, {\it Nonequilibrium Thermodynamics in Biophysics} (Harvard University Press, Cambridge (Mass.), 1965).

\bibitem{19}
B. J. Berne and R. Pecora, {\it Dynamic light scattering} (Wiley, New York, 1976).

\bibitem{20}
D. D. Fitts, {\it Nonequilibrium Thermodynamics} (McGraw-Hill, New York, 1962).

\bibitem{21}
H. Fujita, {\it Foundations of Ultracentrifugal Analysis} (Wiley, New York, 1975).

\end{thebibliography}
\end{document}